\documentclass{article}
\usepackage{graphicx}

\begin{document}

\title{Design Guide for Electronics for Resistive Charge Division in
Thermal Neutron Detection.}


\author{Patrick Van Esch \\
Thomas Gahl \\
Bruno Gu\'erard}

\maketitle

\begin{abstract}
An amplifier has been designed for optimal use of position
sensitive thermal neutron detectors using the principle of
resistive charge division.  The important points in this
optimization are: high counting rates and good spatial resolution.
This amplifier is built as a hybrid circuit and is now used on
several new instruments at the ILL. It consists of a fast low
noise current pre-amplifier, a gaussian shaping circuit based on a
4th order active filter and an essentially noiseless baseline
reconstruction.  In this paper, we present a rather complete
theoretical analysis of the problem that lead us to the choices
made above, and allows for an optimal adaptation to other
situations.  An analysis of unwanted, secondary effects is also
worked out.
\end{abstract}

\section{Introduction.}

The original motivation for the design of this electronics was
given by a project called SANS-2MHZ, in the framework of the
Millennium program at the Institut Laue Langevin, the large
instrument renewal effort the ILL is making in order to be
competitive in the 21st century. The aim of the project was to
make a large area neutron detector with a counting rate of at
least \(2 MHz\)  at 10\% dead time correction. One of the
possibilities considered was to have an array of independent, 1
meter long, 1-dim position sensitive proportional detectors. These
detectors are thin (8 mm diameter) and are based upon resistive
charge division. We will present a general analysis of the problem
of spatial resolution as a function of the electronic noise
sources and take this as a guiding principle to explain our
design.

We will also analyze unwanted secondary effects.  Four different
contributions are considered: extra passive resistance, the
presence of a blocking capacitor, the finite input impedance of
the amplifiers and the presence of a certain capacitive load on
both sides.

\section{Noise analysis of the resistive charge division problem.}

Position sensitive detection based upon charge division is a very
old and well-known technique, described in, for example,
\cite{articlechargediv1}, or in \cite{articlechargediv2}. In a
classical publication by Radeka \cite{articleradeka}, it is shown
that the spatial resolution that can be obtained by resistive
charge division along an \(RC\) line is determined by the
temperature \(T\), the capacity \(C\) and the collected charge
\(Q_{s}\) in the detector:
\begin{equation}\label{eq:resolutionradeka}
    \d L / L = \frac{2.54 \sqrt{k T C}}{Q_{s}}
\end{equation}
when using perfect amplifiers and a signal shaping that is optimal
with respect to this problem.  However, the fundamental hypothesis
formulated to obtain that result is that the generated charge
along the electrode is fast compared to the time scale of \(R C\).
For thermal neutron detection in cylindrical position sensitive
detectors, using a wire and He-3 gas conversion, this hypothesis
is not satisfied at all.  Indeed, a typical detector has, say, \(6
K\Omega / m\) resistance and \(10 pF/m\) capacity.  The total
charge collection time is not negligible compared to the time
constant, which is about 60 ns ; on the contrary, these 60 ns are
almost insignificant compared to the charge collection time. This
charge collection time is of the order of hundreds of nanoseconds,
and is essentially determined by the difference in time of arrival
between the first and the last primary charges along the tritium
and proton tracks arriving in the amplification region on one
side, and the finite drift time of the avalanche ions in the high
field region, which generates the current signal, on the other
side.

Anticipating an (evident) result which will follow in our
analysis, the spatial resolution will improve with a higher signal
(collected total charge) level ; so we will have to work with
shaping functions which are broad (in the time domain) compared to
the time constant of the RC line in order to collect most of the
useful signal. This, on one hand, invalidates the result in
equation \ref{eq:resolutionradeka} in our case, but on the other
hand, simplifies strongly the analysis ; indeed, we do not have to
consider the dynamic behavior of an RC line, with all its Bessel
responses and so on \cite{articleschneider}.  In the low-frequency
limit where we are have to work in order to collect our charges,
the detector essentially behaves as a lumped resistor.

The preamplifier noise contribution can always be modelled as a
series voltage noise source and a parallel current noise source.
The voltage \(v(t)\) and the current \(i(t)\) of both sources are
random processes which are supposed to be stationary, and
gaussian.  They are hence described by their spectral noise
densities.  They could be correlated because they are obtained as
equivalent sources of the noise sources internal to the
preamplifier, but this correlation will be neglected.  When the
first stage of the preamplifier consists of an operational
amplifier, the constructor of the circuit gives those spectra in
the data sheets.  For all practical purposes, they are a constant
(it is white noise).  Only in the case of a MOSFET entry, the 1/f
noise could have a significant contribution.  We will suppose in a
first approach that the amplifier is perfect if we put the noise
sources outside.

\begin{figure}
  \includegraphics[width=10cm]{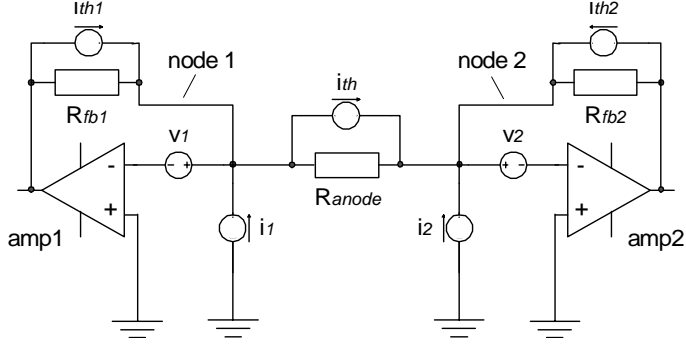}\\
  \caption{Equivalent diagram describing the noise sources in the
  charge division problem in the ideal case.}\label{fig:equivalentnoise}
\end{figure}

The final element which will complete our noise analysis is a
feedback resistor going from the output to the input.  The Johnson
noise of that resistor will add quadratically to the parallel
noise source of the amplifier.  We have now completed the
description of our basic noise circuit in resistive charge
division which is shown in figure \ref{fig:equivalentnoise}.
Although the amplifiers are modelled as operational amplifiers
(and in our case, we really do use such amplifiers), our analysis
applies to just any feedback amplifier structure.
All sources are considered to be independent of each other.  The
measured quantities are the currents in \(R_{fb1}\) and
\(R_{fb2}\). We consider the amplifiers amp1 and amp2 to be
perfect.  Node 1 is hence kept at \(v_1\) and node 2 at \(v_2\).
Solving for the current in the feedback resistors, we have:
\begin{eqnarray}
 \nonumber
  I_{1} &=& i_{th_{1}} + i_{1}- i_{th}+ \frac{v_{2}-v_{1}}{R_{\textrm{anode}}} \\
  I_{2} &=& i_{th_{2}} + i_{2}+ i_{th}+ \frac{v_{1}-v_{2}}{R_{\textrm{anode}}}
\end{eqnarray}

 Let us consider that \(i_{1}\) includes \(i_{th_{1}}\) and that \(i_{2}\)
includes \(i_{th_{2}}\).
When calculating the sum and the
difference of both signals:
\begin{eqnarray}
  I_{-} &=& I_{1}-I_{2} = i_{1} - i_{2} - 2 i_{th} -
  2 \frac{(v_{2}-v_{1})}{R_{\textrm{anode}}} \\
  I_{+} &=& I_{1}+I_{2} = i_{1} + i_{2}
\end{eqnarray}
The spectral densities of the signals combine in square and the
cross terms drop out, so we find, using the Johnson noise
expression for the resistor noise sources:
\begin{eqnarray}
  I_{-}^{2} &=& 2 i_{\textrm{noise}}^{2} + \frac{16 k T}{R_{\textrm{anode}}}+
  8\frac{v_{\textrm{noise}}^{2}}{R_{\textrm{anode}}^{2}} \\
  I_{+}^2 &=& 2 i_{\textrm{noise}}^{2}
\end{eqnarray}
Here, it is understood that \(i_{\textrm{noise}}^{2} =
i_{\textrm{amp}}^{2} + 4 k T/R_{\textrm{fb}}\),
\(i_{\textrm{amp}}\) being the equivalent current noise density of
the amplifier, and \(v_{\textrm{noise}}\) is the equivalent
voltage noise density of the amplifier.   Although these
expressions are in general spectral quantities, in the adopted
hypothesis of white noise, they reduce to numbers.
 In order for the
thermal noise to be the dominant contribution, \(R_{anode}\)
should be in the range:
\begin{equation}
    \frac{v_{\textrm{noise}}^2}{2kT}\ll R_{\textrm{anode}}
    \ll \frac{8 kT}{i_{\textrm{noise}}^2}
\end{equation}
If we take as an example an amplifier and feedback current noise
of \( 2.0 pA/\sqrt{Hz}\) and a voltage noise of \(1
nV/\sqrt{Hz}\), this interval goes from \(125 \Omega\) to
\(8K\Omega\).

If the time-invariant transfer function from input current to
output voltage is \(H(f)\), then the root mean square (r.m.s.) of
the output noise  is given by:
\begin{equation}
    v_{\textrm{r.m.s.}}^{2} = \int_{f=0}^{\infty} i_{n}^{2} |H(f)|^{2} df
\end{equation}
In our case of white noise, this reduces to:
\begin{equation}\label{eq:vrmsfromwhitedensity}
    v_{\textrm{r.m.s.}} =i_{n} \sqrt{\int_{f=0}^{\infty}  |H(f)|^{2} df}
\end{equation}
So knowing the input current noise spectral density, we have to
multiply by a number which depends only on the overall transfer
function (shaping function) to find the 1 sigma deviation on the
output voltage.  Of course, this operation also applies to the sum
and the difference signals. The contribution of the electronic
noise on the spatial resolution of the detector can be determined
as follows.  Consider the dimensionless parameter \(p =
(A-B)/(A+B)\). Here, A and B are instantaneous samples of the
output signals of the amplifiers at their peaking value.  In the
ideal case, p runs from -1 to 1 when the physical position runs
from 0 to length L. The error on p due to the noise, considering
that the noise on A-B and A+B is not correlated (as can easily be
verified assuming identical transfer functions), is given by:
\begin{equation}
    \delta_{\textrm{rms}} p =\frac{ \sqrt{ p^{2} v_{+ \textrm{r.m.s.}}^{2} +
    v_{- \textrm{r.m.s.}}^{2}}}{A+B}
\end{equation}
This translates in a FWHM position resolution of:
\begin{equation}\label{eq:fwhmresolutionfromrmsnoise}
    \delta_{\textrm{FWHM}} x =\sqrt{2 \ln 2} L
    \frac{ \sqrt{ p^{2} v_{+ \textrm{r.m.s.}}^{2} +
    v_{- \textrm{r.m.s.}}^{2}}}{v_{\textrm{signal}}}
\end{equation}
Here, \(v_{\textrm{signal}}\) is the sum of both voltage signals
for an "average" event.

Let us now analyze the general behavior of the influence of the
shaping function on the spatial resolution.   Imagine an initial
charge q, which is deposited instantaneously at the entrance of
the amplifier, so \(i(t) = q \delta(t)\).  The output of the
amplifier is nothing else but the impulse response of the system
times q, and H(f) is the Fourier transform of that impulse
response.  Of course, an overall amplification factor doesn't
influence the spatial resolution as the signal and the noise are
multiplied by it, so in order to compare transfer functions, one
should normalize them such that the peak value (the maximum of the
impulse response) is always equal to 1.  It is easy to work out
that the amplitude normalized impulse response scales in time as:
\begin{equation}
    h_{\tau}(t) = h_{1}(t/\tau)
\end{equation}
and:
\begin{equation}\label{eq:timescalef}
    H_{\tau}(f) = \tau H_{1}(\tau f)
\end{equation}
In these expressions, the subscript 1 corresponds to a
time-normalized impulse response, and \(\tau\) corresponds to the
time scale parameter (the shaping time constant).

The input noise calculations are independent of the transfer
function, and are related to the output noise by the factor given
in equation \ref{eq:vrmsfromwhitedensity}. For an amplitude
normalized transfer function, this factor then gives the absolute
"noise performance" of a shaping function, so we will call this
factor the absolute noise performance factor (ANPF) of the shaping
function.  It should be as low as possible.  It is easy to
determine that the ANPF scales as \(\sqrt{\tau}\) when using the
expression \ref{eq:timescalef}.
 So we have the at first
sight remarkable conclusion that \emph{the faster the amplifier
(the shorter the shaping time) the better the position
resolution}.

 If we can define a time scale parameter which is independent of the
shape of the impulse response, then we can define a 'relative
noise performance factor' by dividing its ANPF by \(\sqrt{\tau}\).
We then have a quantity which can compare different shapes of
transfer functions according to their merit.  The problem with
this definition is of course that a judicious choice of time scale
parameter can artificially give an advantage to one or another
shape so one should be prudent in drawing conclusions.

We can define the gain G of the amplifier as the maximum of the
impulse response.   Writing \(H(f)\) the normalized transfer
function, the full transfer function of an amplifier is then \(G
H(f)\).  In that case, \(v_{\textrm{signal}}\) in formula
\ref{eq:fwhmresolutionfromrmsnoise} is equal to \(G Q_{s} \).  As
expected, this factor G drops out of the expression for the
resolution.
\begin{equation}\label{eq:absoluteresolution}
    \delta_{\textrm{FWHM}} x = \sqrt{2 \ln 2} L \sqrt{\tau} \textrm{RNPF}
    \frac{\sqrt{2 i_{\textrm{noise}}^2 (1 + p^2)
    + \frac{16 k T}{R_{\textrm{anode}}}+ \frac{8
    v_{\textrm{noise}}^2}{R_{\textrm{anode}}^{2}}}}{Q_{s}}
\end{equation}
 Expression \ref{eq:absoluteresolution}  expresses the
resolution due to electronic noise as a function of the shape and
time scale of the overall transfer function, the event charge
produced by the detector, the anode resistance and the equivalent
voltage and current noise sources of the preamplifier.  The
hypothesis is put forward of an otherwise ideal amplifier and full
dynamics. If the sensitive region occupies only a fraction D
\((0<D<1)\) of the whole dynamics of the quantity
\(p=(A-B)/(A+B)\), then the resolution will have to be scaled up
with an extra factor equal to 1/D, see section
\ref{sec:secondaryeffects}.  As long as D is not too far from 1,
we can neglect the influence of these extra components in the
equivalent circuit on the noise calculation.

\section{Gaussian Shaping Function.}

Although the pure pole approximation to a gaussian filter is a
well-known item, we shortly sketch its derivation in order to make
clear all our normalization conventions. The ideal gaussian shape
as an impulse response would be proportional to:
\begin{equation}
    h_{\textrm{perfect}}(t) = \frac{\exp(-t^2/2)}{\sqrt{2 \pi}}
\end{equation}
 its associated transfer
function (in the variable \( \omega = 2 \pi f\) ) equals
\(\exp(-\omega^{2}/2)\), or, in the Laplace variable s,
\(\exp(s^{2}/2)\). Of course, as such, it is not realizable. Using
the well-known approximation scheme \cite{booknetworksynthesis} of
this transfer function into a stable, pure-pole filter by
expanding \(D(s) =1/(H(s) H(-s))\) as a polynomial up to order \(2
n\), and constructing \(H(s)=1/D_{+}(s)\), where \(D_{+}(s)\) is
the polynomial of order n with the same zeros as \(D(s)\) in the
left half complex s-plane, we find for n = 4:
\begin{equation}
    1/D_{+}(s) = \frac{4.90...}{4.90... + 11.42... s + 10.86... s^{2}
    + 5.07... s^3 + s^4}
\end{equation}
where we took the value of the numerator such that H(0) = 1
(because in the process, the amplitude normalization became
arbitrary). The impulse response (the inverse Laplace transform of
H(s)) approaches the perfect gaussian response (up to a time
shift) and corresponds to a realizable, stable pure pole filter of
4th order. We arbitrarily define its time scale \(\tau\) to be
equal to \(2 \pi\), which corresponds to about the time where the
response is "significantly different from 0" (eg, the total
visible width of the pulse).  \(1/\tau\) also corresponds closely
to the cut-off frequency of this low pass filter. The final thing
to do is to normalize the amplitude of the impulse response to 1.
The peaking time of the impulse response occurs at \(0.3114
\tau\), and it turns out that we have to divide the above transfer
function by 2.507 in order to obtain this amplitude normalization.
So we obtain finally:
\begin{eqnarray}
  h_{\tau}(t) &=& 3.86 e^{-8.52 t/\tau}\cos\frac{2.06
    t}{\tau} - 3.86 e^{-7.42 t / \tau}\cos\frac{6.66 t}{\tau} \\
    \nonumber
   &+& 35.33 e^{-8.52 t / \tau}\sin\frac{2.06 t}{\tau}-10.29
    e^{-7.42 t/\tau}\sin\frac{6.66 t}{\tau}
\end{eqnarray}
The corresponding transfer function is:
\begin{equation} \label{eq:gauss4thtransferfunction}
H_{\tau}(s) = \frac{3046.1 \tau}{7635.2 + 2833.4 s \tau + 429.0
s^2 \tau^2 + 31.9 s^3 \tau^3 + s^4 \tau^4}
\end{equation}
Now that we have the normalized transfer function of this filter,
we can calculate the ANPF (by using \(H(s = i 2 \pi f) \) ) and it
turns out to be:
\begin{equation}
    ANPF = 0.3788 \sqrt{\tau}
\end{equation}
Filling this into the expression for the spatial resolution, we
obtain:
\begin{equation}\label{eq:gaussresolution}
    \delta_{\textrm{FWHM}} x = 0.446 L \sqrt{\tau}  \frac{\sqrt{2 i_{\textrm{noise}}^2
    (1 + p^2)
    + \frac{16 k T}{R_{\textrm{anode}}}+ \frac{8
    v_{\textrm{noise}}^2}{R_{\textrm{anode}}^{2}}}}{Q_{s}}
\end{equation}
Let us take as an example a \(1 K\Omega\) resistance, and the same
noise sources as cited earlier.  If we have a detector that
delivers a harmonic average of \(0.4 pC \) per event and we apply
a gaussian shaping time of \(1.4 \mu s\), then on a 1 m long
detector we should be able to achieve a resolution of 12 mm along
the detector.  If the resistance is \(6.5 K\Omega\), the
resolution in the middle of the detector is 5.6 mm, and on the
borders 6.7 mm.  The numbers in this last example have not been
chosen at random: they correspond to an experimental situation we
have set up.  Measurements using a narrow neutron beam on a
prototype Reuter Stokes detector with an overall resistance of
\(6.5 K\Omega\) lead to an observed FWHM resolution 5.7 mm in the
middle of the detector and a deterioration towards 7 mm near the
ends of the detector, confirming our calculation.  In this
measurement, wide word lengths (10 bits) were used to code the
position.  If we measure the r.m.s. noise of the sum and
difference output signals of gaussian amplifiers having a gain of
\(3V/pC\) in this setup, we measure \(6.0 mV_{rms}\) for the
difference and \(4.1 mV_{rms}\) for the sum signal. Calculations
give \(5.7 mV_{rms}\) for the difference signal, and \(3.8
mV_{rms}\) for the sum signal, showing an agreement with the
measurements with less than 10\% deviation.

Let us compare this gaussian transfer function with the "standard"
RC-CR function: its amplitude normalized impulse response is given
by :
\begin{equation}
    h_{rccr}(t) = \frac{t}{\tau}e^{1-\frac{t}{\tau}}
\end{equation}
where \(\tau\) is its peaking time and its transfer function
equals:
\begin{equation}
    H_{rccr}(s) = \frac{e}{\tau (1/\tau + s)^2}
\end{equation}
The ANPF of this RC-CR filter equals \(e/(2\sqrt{2}) \sqrt{\tau} =
0.961 \sqrt{\tau}\). However, in order to really compare the ANPF
values, we should adapt the \(\tau\) in this definition to the one
we used in the gaussian filter.  We will propose identical peaking
times: in that case, we have to replace \(\tau\) by
\(\tau=0.311441 \tau'\) in the above ANPF expression, leading to
\(ANPF = 0.536 \sqrt{\tau'}\).  So we observe that \emph{the
gaussian filter, for identical peaking times, improves the spatial
resolution by about 42\% over the result one would obtain using
the standard RC-CR filter.}

If we take a rectangular shaping function with width \(\tau\),
\begin{equation}
    h_{\textrm{square}}(t) = 1 - \textrm{step}(t-\tau)
\end{equation}
the transfer function is:
\begin{equation}
    H_{\textrm{square}}(s) = \frac{1-e^{-\tau s}}{s}
\end{equation}
and the ANPF equals \(\sqrt{\tau/2}=0.707 \sqrt{\tau} \). Again,
in order to compare the \(\tau\) values, we take the width of the
square function to be the FWHM, the peaking time not being
applicable in this case, of the gaussian approximation which
equals \(0.38171 \tau\). So we should fill in \(\tau = 0.38171
\tau'\) in the expression of the ANPF, to obtain that the \(ANPF =
0.4369 \sqrt{\tau'}\). Again, \emph{the gaussian filter improves
the spatial resolution by 15\% over the result one would obtain
with a square filter for identical FWHM times}.

All this indicates that the gaussian transfer function as derived
above is a very acceptable solution to the shaping problem for
resistive charge division.

\section{The electronic implementation.}

The analogue part of the circuit implemented at the I.L.L. in the
frame of the SANS-2MHZ program takes on the following form: a
pre-amplifier which acts as a current amplifier, a gaussian filter
implemented as an active filter and a baseline correction circuit.

\subsection{The preamplifier}

 The preamplifier needs to satisfy 3
criteria: the amplifier noise contributions have to be such that
they don't deteriorate significantly the spatial resolution, the
input impedance has to be as low as possible and it has to stand
very high counting rates. The fact that there is anyway a
resistive noise source (the resistive anode) eliminates the noise
advantages a charge amplifier has. Indeed, in order to stand very
high counting rates, we opted for a current amplifier, which
suffers no pile up effects. The transfer function of the circuit
is chosen to be a first order system with a time constant of the
order of the time constant of the detector, about 60 ns.  In order
to achieve a very low input impedance, a high open loop gain is
required, as explained in subsection \ref{subsec:impedance}. The
high bandwidth combined with the high open loop gain made us
decide to go for a bipolar operational amplifier.

Expression \ref{eq:gaussresolution} is the guiding principle to
select the operational amplifier, which has to have a very high
open loop unity gain bandwidth.  The best values one can obtain
for the current noise are around \(1.5 pA/\sqrt{Hz}\), which
doesn't contribute significantly to the resolution as long as the
anode resistance is below \(\sim 10 K\Omega\). This comforts our
choice of a bipolar circuit, such as the CLC425 from National
Semiconductor, unfortunately now not in production anymore. The
feedback resistor doesn't have a significant influence compared to
the current noise of the amplifier as long as it is above about
\(12 K\Omega \). A feedback capacitor of about 5 pF lets us obtain
the right bandwidth in this case, which is a satisfying value.

\begin{figure}
  \includegraphics[width=10cm]{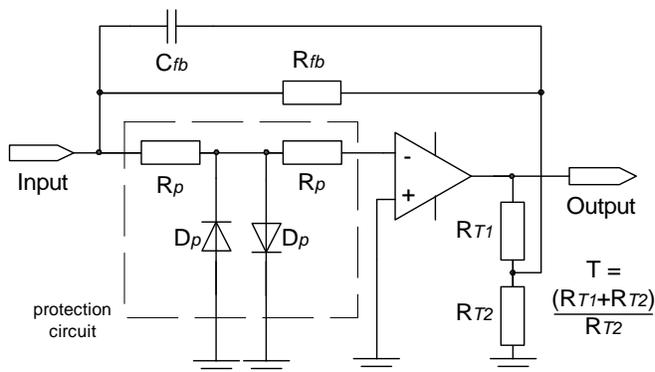}\\
  \caption{Basic scheme of the preamplifier.}\label{fig:preamplifier}
\end{figure}

 The
basic scheme that we obtain for the structure of the preamplifier
is shown in figure \ref{fig:preamplifier}.

As long as the voltage noise is a few \(nV/\sqrt{Hz}\), and the
anode resistance is not too low (above \(\sim 1 K\Omega\)) this
term doesn't contribute significantly to the spatial resolution.
 If a resistive protection circuit against
accidental discharges is required, care should be taken to include
it in the feedback loop (as shown in figure
\ref{fig:preamplifier}), otherwise its resistance adds to the
input impedance. One should be warned, however, that the resistive
part of such a circuit will add to the voltage noise of the
circuit.

In order to obtain a larger gain, a T-type feedback structure can
be used (see figure \ref{fig:preamplifier}). An extra advantage of
using a current amplifier over a charge amplifier is that no
pole-zero cancellation network is needed. The preamplifier is AC
coupled to the gaussian filter.

\subsection{The gaussian filter.}

The fact that for neutron detection, time constants of the order
of the microsecond are used allows us to use operational
amplifiers (designed for the telecom industry) in active filters
to implement the gaussian transfer function. Factorizing
expression \ref{eq:gauss4thtransferfunction} into two second-order
contributions, we notice that the Q-factors of both circuits are
quite low and that hence a simple Multiple Feedback Network
\cite{bookactivefilters} can be used to implement them. Although
possible, passive LC filters require rather large self inductance
values on this time scale, so we reserve that kind of filters to
faster circuits.

\subsection{The baseline correction circuit.}

\begin{figure}
  \includegraphics[width=10cm]{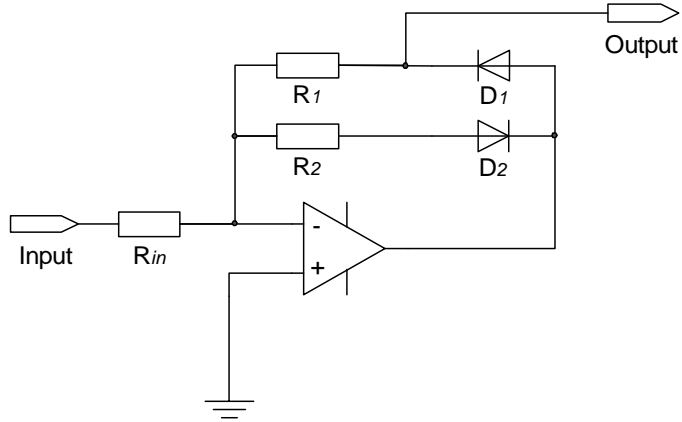}\\
  \caption{The baseline correction circuit: extraction of the baseline.}
  \label{fig:baselinecorrection}
\end{figure}

 Given the fact that the
gaussian impulse response is unipolar, there is a problem of a
shifting baseline at high counting rates.  A simple clamping
circuit has the disadvantage of clamping on the noise, so we would
have a noisy baseline correction.  In order to solve this problem,
we use the circuit displayed in figure
\ref{fig:baselinecorrection}.
In that figure, \(R_{\textrm{in}}\) translates the input voltage
into a current. If the voltage is positive, the feedback loop is
closed through resistor R2 and diode D2 ; the voltage at the
output is 0. If the input voltage is negative, the loop is closed
through R1 and D1. The output voltage is now \(-
R1/R_{\textrm{in}} V_{\textrm{in}}\), which is a positive
quantity. Taking \(R_{\textrm{in}} = R1\), if the signal from the
gaussian shaper is applied to this circuit, then the output equals
minus the baseline which is negative.  It is now sufficient to low
pass filter this output and add it to the input signal with a
simple active adder circuit to restore the baseline.  The path
R2/D2 prevents the feedback loop from ever opening completely, and
hence keeps the operational amplifier from saturating. If the
application is continuous irradiation, then the baseline will vary
very smoothly and a simple first order low pass (RC) circuit with
a cutoff frequency below say \(1 KHz\) can do. However, in pulsed
time of flight applications, the baseline also fluctuates more
rapidly, and a more sophisticated filter might be necessary in
order to filter all noise and glitches, but reacting fast enough
to the changing baseline.

\subsection{The conversion circuitry.}

An analogue peak detector with threshold on the sum signal
triggers 2 single shot 12 bit ADC converters with a dynamic range
of 5 V. The 11 most significant bits of each conversion are used
to compose a 22 bit word which serves as an address in a position
lookup table. This lookup table is materialized by an EPROM
circuit ; the result of the lookup operation is the position of
the impact in an 8-bit word. The quantization noise of the ADC (\(
= 5V / 2^{11} 1/\sqrt{12} = 0.7 mV_{\textrm{rms}} \) ) should be
smaller than the noise at the output of the amplifiers.  It should
be noted that here, the right choice of the gain to map the signal
range to the dynamic range, is important. The quantization noise
added to the resolution by truncating the result to an 8 bit word
results in a FWHM resolution of \( L/2^8 1/\sqrt{12} \sqrt{8 \ln 2
} = 0.00266 L\), which should be added in square to the resolution
obtained in \ref{eq:gaussresolution}.  Taking our measurements
again, which had a spatial resolution of 5.7 mm using wide word
lengths, when coding on 8 bits, using the above formula, we should
increase theoretically the resolution to 6.3 mm.  Measurements
lead to 6.6 mm of resolution when coding on 8 bits.

\section{Secondary effects.}

\label{sec:secondaryeffects}

In order to perform the above analysis, we used a "perfect"
circuit.  In reality, there are unavoidable secondary effects
which have an influence on the performance.  We will study the 4
most prominent ones, and make the assumption that the situation
remains symmetrical.  First we will study the presence of an
additional passive resistance.  Interesting in its own right, it
also introduces the notion of "equivalent resistance" for any
other contribution to the reduction of the dynamics, and hence
makes it possible to compare the importance of several different
contributions.  A most important influence is played by the
blocking capacitor, which is a necessary component to protect the
electronics from the high bias potential applied to the resistive
electrode.  It has two effects: one is a reduction of dynamics
(just as the extra resistance), the other is to introduce an
undershoot in the signal.  The finite input impedance of the
preamplifier can also contribute significantly to the loss in
dynamics.  Finally, the presence of a capacitive load (due to the
wiring up of the detector) will not influence the dynamics, but
will introduce a noise contribution that can deteriorate the
spatial resolution.

\begin{figure}
  \includegraphics[width=10cm]{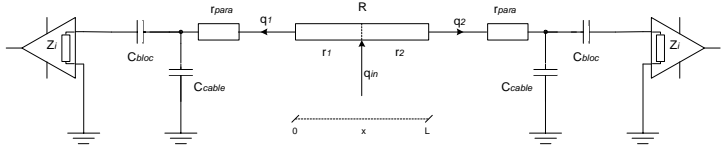}\\
  \caption{Equivalent circuit of all secondary effects we are considering.}
  \label{fig:secondaryeffects}
\end{figure}

We will study all these effects one by one but we will not study
their combined effects. The overall scheme is displayed in figure
\ref{fig:secondaryeffects}.
The ideal case would give us a relationship between the position
\(x\) and the collected charges \(q_1\) and \(q_2\) given as
follows:
\begin{equation}
    q_1 = \left( 1-\frac{x}{L}\right) q_{\textrm{in}} ;
    q_2 =  \frac{x}{L} q_{\textrm{in}}
\end{equation}
We can hence introduce an ideal dimensionless position \(p\) as
follows:
\begin{equation}
 p = \frac{q_2-q_1}{q_1+q_2} = \frac{r_1-r_2}{r_1+r_2}
\end{equation}
When \(p = -1\), we hit position \(x=0\), and when \(p = +1\), we
hit position \(x = L\). We will observe that several effects
change this relationship into:
\begin{equation}
 p = \frac{q_2-q_1}{q_1+q_2} = D \frac{r_1-r_2}{r_1+r_2}
\end{equation}
where \(D\) is a number between 0 and 1, and is called the
dynamical factor.

\subsection{Additional resistance.}

 Imagine that the resistive
electrode continues to be resistive beyond the sensitive area, or
that some resistance is present in the conducting circuit before
reaching the --- supposed perfect --- current amplifier at 0
impedance (virtual ground).  Let us suppose that the case is
symmetrical and that an extra resistance \(r_{\textrm{parasite}}\)
is added at both sides. We have now, for an injection at point x
of a total length L:
\begin{eqnarray}
  q_1 &=& q_{\textrm{in}}\frac{r_2 + r_{\textrm{parasite}}}{R+2r_{\textrm{parasite}}} \\
  q_2 &=& q_{\textrm{in}}\frac{r_1 + r_{\textrm{parasite}}}{R+2r_{\textrm{parasite}}}
\end{eqnarray}
If we now calculate the range that can be taken up by
\(p=(q_2-q_1)/(q_1+q_2)\):
\begin{equation}
    p = \frac{r_1-r_2}{R+2
    r_{\textrm{parasite}}}=\frac{R}{R+2r_{\textrm{parasite}}}\frac{r_1-r_2}{r_1+r_2}
\end{equation}
we observe that the dynamical factor is:
\begin{equation}
    D =\frac{R}{R+2r_{\textrm{parasite}}}\simeq 1 - 2\frac{r_{\textrm{parasite}}}{R}
\end{equation}
So the calibration, \((q_2-q_1)/(q_1+q_2)\) as a function of
\((r_1-r_2)/(r_1+r_2)\), stays linear, but the slope is reduced.
The second approximation is valid if the parasitic resistance is
small compared to the useful resistance of the resistive
electrode.

\subsection{Blocking capacitor.}

If we introduce a blocking capacitor \(C_{\textrm{bloc}}\), or
shortly \(C\), on each side (see figure
\ref{fig:secondaryeffects}), we find as current transfer functions
(injected current to collected current):
\begin{eqnarray}
  i_1(s) &=& \frac{1 + C r_2 s}{2 + C (r_1 + r_2)s} \\
  i_2(s) &=& \frac{1 + C r_1 s}{2 + C (r_1 + r_2)s}
\end{eqnarray}
If we look at the time domain responses of these transfer
functions:
\begin{eqnarray}
i_1(t) &=& \frac{e^{-\frac{2
t}{C(r_1+r_2)}}(r_1-r_2)}{C(r_1+r_2)^2}
+\frac{r_2}{r_1+r_2}\delta(t) \\
i_2(t) &=& \frac{e^{-\frac{2
t}{C(r_1+r_2)}}(r_2-r_1)}{C(r_1+r_2)^2}
+\frac{r_1}{r_1+r_2}\delta(t)
\end{eqnarray}
The time signals clearly contain two parts: the second term is
exactly what we would have if we were in the ideal case, and the
first term is the effect of the capacitors.  It is an exponential
decay with a time constant of \(C (r_1 + r_2)/2\). It has moreover
the opposite amplitude behavior than the ideal term: indeed, when
the injection point goes more to the left, so that the \(i_1\)
current increases, the extra term becomes more and more negative.
On the other hand, when the injection point goes to the right, so
that the ideal term almost vanishes, the extra term becomes
positive and more and more important.

The two extra terms in \(i_1\) and \(i_2\) are opposite, so that
they cancel in the sum.  This is something important: \emph{on the
sum signal, the effect of the extra capacitors vanishes}. This
means that the sum signal contains the timing information,
independent of the position of the impact.

 We should of course convolve this injected current
with the overall impulse response of the shaping amplifier. There
are two contributions:
\begin{eqnarray}
  i_1 \star h(t) &=& (r_1 - r_2) V(t) + \frac{r_2}{r_1+r_2}h(t) \\
 i_2 \star h(t) &=& (r_2 - r_1) V(t) + \frac{r_1}{r_1+r_2}h(t)
 \end{eqnarray}
In these expressions, we introduced the function V(t):
\begin{equation}
    V(t) = \frac{e^{-2\frac{t}{RC} } }{CR^2}\star h(t)
\end{equation}
V(t) doesn't depend on the injection point along the electrode,
but only on \(R = r_1+r_2\). We notice two things: while the
second term in each expression gives exactly what we wanted to
obtain, the first one disturbs the response in a non-proportional
way.  This means that the individual maxima of the two functions
will in general not occur together in time, and the relative
deformation of the signal will depend on the injection point.  If
we would take those individual maxima as charges to be collected,
the calibration curve would be non-linear. However, if we can take
a common sample time \(T\) for both curves which is independent of
the injection point, and which corresponds to the maximum of the
impulse response (G), we find the following:
\begin{eqnarray}
  q_1 &=& (r_1-r_2) V(T) + \frac{r_2}{r_1+r_2} G\\
  q_2 &=& (r_2-r_1) V(T) + \frac{r_1}{r_2+r_1} G
\end{eqnarray}
\begin{equation}
    p = \frac{q_2-q_1}{q_2+q_1}=\frac{r_1-r_2}{r_1+r_2}
    \left(1-\frac{2(r_1+r_2)V(T)}{G}
    \right)
\end{equation}
So we see here that the calibration curve corresponds to a
straight line, with a reduction factor of the dynamics that is
equal to:
\begin{equation}
    D =1-\frac{2RV(T)}{G}= 1 - 2 \frac{1}{GRC}
    \left(e^{-2\frac{t}{RC} } \star
    h(t)\right)_{t=T}
\end{equation}
 When we compare the expression of the loss in
dynamics with the one for a resistor at both ends, we can identify
both formulas in the approximation of small deviation from unity.
In that case, it is as if the coupling capacitor introduced an
equivalent parasitic resistance:
\begin{equation}
    r_{\textrm{eq}} =\frac{1}{GC}
    \left(e^{-2\frac{t}{RC} } \star
    h(t)\right)_{t=T}
\end{equation}
If we assume that the time constant \( CR\) is larger than the
time scale (and of course the sampling time) of the shaping
function, then the exponential with which we have to convolve
looks essentially like a unit step function.  If we replace it in
the convolution integral, we have:
\begin{equation}
    r_{\textrm{eq}} \simeq \frac{1}{GC}\int_{t=0}^{T} h(t) dt
\end{equation}
The equivalent resistance of the coupling capacitor turns out to
be simply inversely proportional to the capacity times a constant
that only depends on the shaping function.  It really has a
meaning of an equivalent resistance.  It turns out that for a
Gaussian 4th order filter function, this expression is numerically
equal to:
\begin{equation}
\label{eq:equicapagauss}
 r_{\textrm{eq}} \simeq  0.1576 \frac{\tau}{C}
\end{equation}
 Here, \(\tau\) is
the  time scale of the Gaussian shaper as we defined it.

 A remark of a technical nature is maybe due: the capacitance is of
course the dynamic capacitance (small signals) for the capacitor
under the static bias load that one has in mind. For a perfect
capacitor, that doesn't make any difference, of course ; however,
for ceramic capacitors using the X7R dielectric, for example, that
dynamic capacity is reduced by 40\% when working at the nominal
voltage !

A second problem, introduced by the presence of blocking
capacitors, is the fact that there can be an undershoot on the
"large" signal.  This undershoot is a problem in the case a
baseline correction is used.  We can estimate an upper boundary of
the value of the undershoot as follows:  given that we are just
outside of the time lapse where the shaper function is important
(if it is a well-peaked function such as a Gaussian), the
undershoot is essentially the sample value of the "non-ideal" part
of the impulse response at that moment.  Of course, there will be
a slight contribution of the shaper response, so the actual
undershoot will be slightly less than this value.  But we will
obtain a reliable upper boundary.  Let us calculate this value now
for \(r_1 = 0\) and \(r_2 = R\):
\begin{equation}
    u_1 \simeq R V(T')
\end{equation}
We sample at \(t = T'\), assuming that h(t) has no contribution
anymore there. This can be worked out as follows:
\begin{equation}
    u_1 \simeq R \frac{e^{-2\frac{t}{RC} } }{CR^2}\star h(t)
    = \frac{1}{RC}e^{-\frac{2}{RC}T' }\int_{t=0}^{T'}h(t)
    e^{\frac{2t}{RC}}dt
\end{equation}
The last integral can be extended to infinity (because h is
supposed to be negligible outside of the range 0-T'). As H(s)
normally has poles with a real value much more negative than the
exponential coefficient, this operation is allowed. We can replace
the integral by a sample point of the Laplace transform of h (the
shaper transfer function).  In that case, we finally obtain as our
undershoot estimation (in absolute value):
\begin{equation}
    u_1 \simeq \frac{1}{RC}e^{-\frac{2}{RC}T' } H\left(s = \frac{-2}{RC} \right)
\end{equation}
To obtain the relative undershoot, we have to divide that
expression by the gain of the shaper function. Knowing the scaling
relations of the transfer function in the case of unity gain, we
see that the undershoot only depends on the ratio of \(T'/(RC)\).
In the case of a gaussian filter, then taking \(T' = \tau\) gives
us numerically:
\begin{equation}
\label{eq:undershootgauss}
    u_1 \simeq \frac{3046.1 x e^{-2 x}}
    {7635.3 + 2833.4 (2x) + 429.01 (2x)^2 + 31.9 (2x)^3 + (2x)^4}\simeq 0.399 x
\end{equation}
where:
\begin{equation}
    x =
    \frac{\tau}{R C}\ll 1
\end{equation}
The undershoot introduces an essentially random error in the case
of a baseline correction, with an amplitude limited to the above
amount. It depends on the image that is projected onto the
detector. It should hence be smaller than the relative error
introduced by the noise over the full dynamics of the signals.

We use 68nF capacitors as blocking capacitors in our application.
With \(1.4 \mu s\) gaussian shaping, this corresponds to an
equivalent resistance of \(3.2 \Omega\) which is really
negligible.  The maximal undershoot, given a dynamics of 5V, and a
detector resistance of \(6.5 K\Omega\), is \(6.3 mV\).  This is of
the same order of magnitude as the noise levels (see our
previously calculated example).  We see here that it is in fact
the undershoot which gives us the most severe condition on the
value of the blocking capacitor.

\subsection{Input impedance.}

\label{subsec:impedance}

 The input impedance of the preamplifier
circuit also has as an effect to reduce the effective dynamics of
the charge division. We will give an estimation of this effect.
The input impedance of a current amplifier with a transimpedance
Z(s) determined by a passive feedback loop, is easily found out to
be equal to:
\begin{equation}
    Z_{in}(s) = \frac{Z(s)}{A(s)}
\end{equation}
where A(s) is the open loop transfer function of the amplifier. We
will take it that Z(s) is represented by a first order system. In
figure \ref{fig:preamplifier}, this is the feedback resistor
\(R_{\textrm{fb}}\) times the gain factor T of the T-network, and
the time constant \(t_{\textrm{pre}}\) is given by the feedback
resistor and its parallel capacitor. It is a bit more delicate to
write down the open loop transfer function A(s) of the operational
amplifier: indeed, large bandwidth amplifiers usually have very
complicated and not very well known transfer functions. However,
because of the gaussian filter, we only use a limited part of that
bandwidth, and we will assume that we can model the open loop gain
by a first order system with dc gain \(A_{ol}\) and time constant
\(t_{ol}\). The impedance then becomes:
\begin{equation}
    Z_{\textrm{in}}(s) = \frac{T_{\textrm{fb}} R_{\textrm{fb}}
    (1+ t_{\textrm{ol}}s)}{A_{\textrm{ol}}(1+t_{\textrm{pre}}s)}
\end{equation}
If we now define the following equivalent elements:
\begin{eqnarray}
  r &=& \frac{T_{\textrm{fb}} R_{\textrm{fb}} }{A_{\textrm{ol}}} \\
  l &=& \frac{T_{\textrm{fb}} R_{\textrm{fb}}
  t_{\textrm{ol}} }{A_{\textrm{ol}}} = t_{\textrm{ol}} r
\end{eqnarray}
we can rewrite the input impedance as:
\begin{equation}
    Z_{\textrm{in}}(s) = \frac{r + l s}{1 + t_{\textrm{pre}} s}
\end{equation}

At first sight, one would be inclined to neglect the pole in this
expression.  Indeed, the preamplifier is supposed to represent a
current amplifier, and hence be much faster than the gaussian
shaping filter that follows.  In that case, the input impedance of
the amplifier is equivalent to the series circuit of a resistor of
value \(r\) and an inductor of value \(l\).  We tried this, and to
our surprise the results are numerically quite different from the
results obtained when we don't neglect this pole.  So this pole
cannot, in the cases we're interested in, be neglected.

\begin{figure}
  \includegraphics[width=10cm]{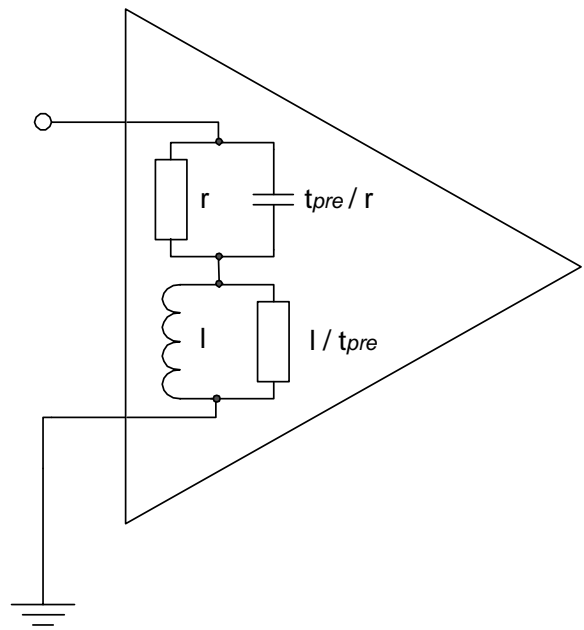}\\
  \caption{The equivalent circuit of the input impedance
  of the amplifier.}\label{fig:inputimpedance}
\end{figure}

If we analyze the structure of the input impedance, we can easily
show \cite{booknetworksynthesis} that this corresponds to the
passive circuit displayed in figure \ref{fig:inputimpedance}.
Using the same approach as in the case of the blocking capacitor,
we find that the currents take on the form:
\begin{eqnarray}
  i_1(t) &=& -\frac{(r_1-r_2)(l - r t_{\textrm{pre}})}{(2 l + (r_1+r_2)
  t_{\textrm{pre}})^2}
  e^{-\frac{2 r + r_1 +r_2}{2 l + (r_1+r_2) t_{\textrm{pre}}}t}
  +\frac{l + r_2 t_{\textrm{pre}}}{2 l + (r_1 + r_2) t_{\textrm{pre}}}\delta(t) \\
i_2(t) &=& \frac{(r_1-r_2)(l - r t_{\textrm{pre}})}{(2 l +
(r_1+r_2) t_{\textrm{pre}})^2}
  e^{-\frac{2 r + r_1 +r_2}{2 l + (r_1+r_2) t_{\textrm{pre}}}t}
  +\frac{l + r_1 t_{\textrm{pre}}}{2 l + (r_1 + r_2) t_{\textrm{pre}}}\delta(t)
  \end{eqnarray}
Again, the sum of both signals is not affected by the presence of
this input impedance.  We next  convolve these expressions with
the transfer impedance of the shaping amplifier, which in our case
is the 4th order gaussian filter function.  Finally, sampling at
the time when the sum signal attains its maximum, we obtain the
sampled charges \(q_1\) and \(q_2\) at each side. The calibration
remains linear, but a reduction of dynamics is introduced by this
input impedance.  This allows us to define an equivalent
resistance. The result of these algebraic operations is:
\begin{equation}
    r_{\textrm{equi}} = R \frac{r R t_{\textrm{pre}} V(\alpha) +
    l (2r + R - R V(\alpha))}
    {R t_{\textrm{pre}}+2 r t_{\textrm{pre}}(1-V(\alpha))+2 l V(\alpha)}
\end{equation}
In this expression, we defined \(R = r_1 + r_2 \) the anode
resistance, and:
\begin{equation}
    \alpha = \frac{2 r + R}{2l + R t_{\textrm{pre}} }\tau_{\textrm{gauss}}
\end{equation}
The function \(V(\alpha) \) is now defined as follows:
\begin{equation}
    V(\alpha) = \alpha e^{(-0.3114 \alpha)}\int_{t=0}^{0.3114}h_1(t)
    e^{\alpha t}dt
\end{equation}
using the time normalized gaussian filter function \(h_1(t)\). We
observe that the equivalent resistance doesn't depend on the
injection point (is no function of \(r_1\) or \(r_2\)
individually), which confirms the linearity of the calibration,
but it is a function of the total anode resistance \(R\). Working
out numerically the integral, we find:
\begin{equation}
    V(\alpha) = \frac{\alpha (3046.1 e^{-0.3114 \alpha}
    +(\alpha - 12.02)(153.3 - 19.9\alpha+\alpha^2))}
    {(76.8-17.0 \alpha + \alpha^2)(99.5-14.8\alpha+\alpha^2)}
\end{equation}

 We can obtain an interesting result
when we take the limit of big R, meaning, much bigger than
\(2l/t_{\textrm{pre}}\) or \(r\). \(\alpha\) then converges to a
limiting value given by:
\begin{equation}
    \alpha_{\textrm{lim}}=\frac{\tau_{\textrm{gauss}}}{t_{\textrm{pre}}}
\end{equation}
and the expression for the equivalent resistance reduces to:
\begin{equation}
    r_{\textrm{equi}}^{\textrm{lim}}=\frac{l(1-V(\alpha_{\textrm{lim}}))}{t_{\textrm{pre}}}
    +rV(\alpha_{\textrm{lim}})
\end{equation}
We now observe that for large enough detector resistance R, the
equivalent input resistance of the circuit is a well-defined
quantity. In cases where the input impedance is a crucial
parameter, we should optimize the ratio of the gaussian shaper and
the preamplifier time constant such that the associated equivalent
resistance is minimized.  In fact,  the perfect match occurs  when
the preamplifier time constant is equal to
\(l/r=t_{\textrm{ol}}\). Usually, this time constant is too big to
be used (it should still be smaller than the gaussian shaping
time), so a less optimal choice has to be made. Diminishing the
gain of the preamplifier also reduces the input impedance.

As an example, we have taken a preamplifier based on a CLC425
operational amplifier (modelled with open loop dc gain of 20000
and corner frequency \(100 kHz\)), a feedback resistor of \(12
K\Omega\) in parallel with a \(5.6 pF\) feedback capacitor, and a
T-factor of 13.0.  With a \(1.4 \mu s\) gaussian shaping, the
above calculation leads to an equivalent resistance of \(26.9
\Omega\) when the detector is \(500 \Omega\) ; we have measured
\(26 \Omega\).  The limiting value is \(21.0 \Omega\).

\subsection{Capacitive load.}

We will now analyze the effect of an extra load capacitor on both
sides of the detector, as displayed in figure
\ref{fig:secondaryeffects}.  This corresponds partly to the
detector capacity itself, but mainly to the extra capacitive load
that is introduced by the conductors leading from the sensitive
electrode up to the input of the amplifiers.  An order of
magnitude is 1 pF per cm of (coaxial) connection wire.  If we add
this capacity C on both sides of the noise equivalent network in
figure \ref{fig:equivalentnoise}, the noise currents now become:
\begin{eqnarray}
 \nonumber
  I_{1} &=& i_{th_{1}} + i_{1}- i_{th}+ \frac{v_{2}-v_{1}}{R_{\textrm{anode}}}
  - s C v_1 \\
  I_{2} &=& i_{th_{2}} + i_{2}+ i_{th}+ \frac{v_{1}-v_{2}}{R_{\textrm{anode}}}
  - s C v_2
\end{eqnarray}
The spectral densities of the sum and difference signals then take
on the following form:
\begin{eqnarray}
  I_{-}^2(f) &=& 2 i_{\textrm{noise}}^2+\frac{16 k T}{R_{\textrm{anode}}}+
  8v_{\textrm{noise}}^2(\pi^2 f^2 C^2 +\frac{1}{R_{\textrm{anode}}^2}) \\
  I_{+}^2(f) &=& 2 i_{\textrm{noise}}^2 +
  8v_{\textrm{noise}}^2\pi^2 f^2 C^2
\end{eqnarray}
So to both densities, we add the quantity:
\begin{equation}
i_C^2(f) = 8v_{\textrm{noise}}^2\pi^2 f^2 C^2
\end{equation}
 This time it is clear that these noise spectral densities are
a function of frequency f.  We cannot apply the simple factor
anymore as in equation \ref{eq:vrmsfromwhitedensity}.  In fact,
our noise formulas are of the form: \(a+ b f^2 \), so we can still
use \ref{eq:vrmsfromwhitedensity} for the a-part, but we now need
a second number for the b-part.  The extra part equals:
\begin{equation}
    v_{\textrm{r.m.s.}-C}^{2} = 8 \pi^2 v_{\textrm{noise}}^2 C^2
    \int_{f=0}^{\infty} f^2 |H(f)|^{2} df
\end{equation}
We define now the capacitive noise number \(\eta\):
\begin{equation}
    \eta = \int_{f=0}^{\infty} f^2 |H(f)|^{2} df
\end{equation}
It can easily be verified that \(\eta\) scales as \(1/\tau\), and
not, as is the case for the white noise contribution, as \(\tau\).
So this capacitive load noise becomes more and more important when
the time scale becomes small.  We can call \(\sqrt{\eta}\) the
capacitive noise performance factor (CNPF) if we work with the
amplitude normalized transfer function. For the gaussian 4th order
approximation, we can work out that:
\begin{equation}
    \textrm{CNPF} = \frac{0.2856} {\sqrt{\tau}}
\end{equation}
This means that we have an effective extra contribution to the sum
and difference noise current densities (considered as white
noise), equal to:
\begin{equation}
    i_C^{\textrm{eff}}= \frac{2 \sqrt{2}\pi \textrm{CNPF} v_{\textrm{noise}} C}
    {\textrm{ANPF}}
\end{equation}
For the gaussian shaping function, this turns out to be:
\begin{equation}
\label{eq:extracapanoisegauss}
    i_C^{\textrm{eff}} = 6.70 \frac{v_{\textrm{noise}} C }{\tau}
\end{equation}
As an example, a \(1.0 nV/\sqrt{Hz}\) voltage noise density over a
capacity of 100 pF (1 meter of coaxial cable) and a time constant
of \(1\mu s\) gives us an effective noise current of \( 0.67
pA/\sqrt{Hz}\).  This is still smaller than the other noise
current contributions (but its effect should start to become
visible).

In order to add this effect to the spatial resolution, it is
sufficient to replace the sum and difference currents by their
combination, squared, with this effective current.

\section{Conclusion.}

 The principles leading to the design of the
front end electronics for resistive charge division in thermal
neutron detection in high counting rate applications are exposed.
The choice we made for the analogue part consists of three sub
functions: a preamplifier which acts as a current amplifier, based
upon a fast, bipolar operational amplifier ; a 4th order gaussian
shaping circuit based upon simple active filter circuits, and a
noiseless baseline correction.  A detailed analysis is presented
leading to the expression (equation \ref{eq:absoluteresolution})
which gives the spatial resolution as a function of the detector
signal, the noise sources of the preamplifier and the gaussian
shaping time constant.  A comparison of the gaussian transfer
function with the standard RC-CR and the square pulse serves as a
motivation for the gaussian shape.  A short description of the
practical implementation as hybrid circuits follows.

A profound analysis of several secondary effects are given.  Two
important effects are due to the blocking capacitor.  Its
equivalent dynamical resistance is given in equation
\ref{eq:equicapagauss},
 and an upper boundary to the undershoot it introduces is
 given in equation \ref{eq:undershootgauss}.
 Also a detailed analysis of the influence of the input impedance
of the amplifier is presented, leading to indications how to
optimize this quantity if necessary.  Finally, the noise
contribution due to the capacitive load is worked out in equation
\ref{eq:extracapanoisegauss}.

Some experimental results confirm the calculations presented here.

Although our analysis was focussed on the design of the front end
electronics for the SANS-2MHz project at the Institut Laue
Langevin, the results are of a sufficiently general nature to
allow them to be a guide for other designs in position sensitive
detection.

\end{document}